\documentclass[12pt,showpacs,preprintnumbers,superscriptaddress,nofootinbib]{revtex4}
\usepackage{amssymb}
\usepackage{amsmath, graphicx}
\usepackage{dcolumn}
\usepackage{bm}
\usepackage{epstopdf}
\usepackage[utf8]{inputenc}
\usepackage{color}

\newcommand{\hs}{\hspace*{0.3cm}}

\newcommand{\be}{\begin{equation}}
\newcommand{\ee}{\end{equation}}
\newcommand{\bea}{\begin{eqnarray}}
\newcommand{\eea}{\end{eqnarray}}
\newcommand{\ben}{\begin{enumerate}}
\newcommand{\een}{\end{enumerate}}
\newcommand{\bde}{\begin{widetext}}
\newcommand{\ede}{\end{widetext}}
\newcommand{\nn}{\nonumber}
\newcommand{\crn}{\nonumber \\}

\newcommand{\fr}{\frac}

\newcommand{\bc}{\begin{center}}
\newcommand{\ec}{\end{center}}

\begin{document}

\title{\boldmath Baryogenesis in the Zee-Babu model with arbitrary $\xi$ gauge}
\author{Vo Quoc Phong}
\email{vqphong@hcmus.edu.vn}
\affiliation{ Department of Theoretical Physics, VNUHCM-University of Science, Ho Chi Minh City,Vietnam}
\author{Nguyen Chi Thao}
\email{ncthao@grad.iop.vast.ac.vn}
\affiliation{ Graduate University of Science and Technology, Vietnam Academy of Science and Technology, 18 Hoang Quoc Viet, Cau Giay, Hanoi, Vietnam}
\affiliation{Institute of Physics, Vietnam Academy of Science and Technology, 10 Dao Tan, Ba Dinh, Hanoi, Vietnam}	
\author{Hoang Ngoc Long}
\email{hoangngoclong@tdt.edu.vn}
 \affiliation{Theoretical Particle Physics and Cosmology Research Group, Advanced Institute of Materials Science,  Ton Duc Thang University, Ho Chi Minh City, Vietnam}
 \affiliation{Faculty of Applied Sciences,
Ton Duc Thang University, Ho Chi Minh City, Vietnam}
\date{\today }

\begin{abstract}
		We consider the baryogenesis picture in the Zee-Babu model. Our analysis shows that electroweak phase transition (EWPT) in the model is a first-order phase transition at the $100$ GeV scale, its strength ranges from 1 to 4.15
  and the masses of charged Higgs boson  are smaller than $300$ GeV. The EWPT is strengthened by only the new bosons and this strength is enhanced by arbitrary $\xi$ gauge. However, the $\xi$ gauge does not break the first-order EWPT or, in other words, the $\xi$ gauge is not the cause of the EWPT. This  leads to the fact  that the calculation of EWPT in Landau gauge is enough;
and the latter may provide baryon-number violation (B-violation) necessary for baryogenesis in the relationship with nonequilibrium physics in the early universe.	
	\end{abstract}
	\maketitle

\section{INTRODUCTION}\label{secInt}

Physics, at present, has entered into a new period, on the understanding
 the early Universe. In that context,  Cosmology and Particle Physics are on
the same way. Being as a central issue of cosmology and particle physics, at present the baryon asymmetry is an interesting problem.
If we could explain this problem, we can understand the true nature of the smallest elements and reveal a lot about an imbalances
 matter-antimatter from the early Universe.

The electroweak baryogenesis (EWBG) is a way to explaining the baryon asymmetry of the Universe (BAU) in the early Universe, associating with
 Sakharov conditions, which are B, C, $CP$ violations, and deviation from thermal equilibrium  \cite{sakharov}.
 These conditions can be satisfied when the EWPT must be a strongly first-order phase transition. Because that not only
  leads to thermal imbalance  \cite{mkn}, but also makes a connection between B and $CP$ violation via nonequilibrium physics \cite{ckn}.

The EWPT has been investigated in the standard model (SM) Ref.\cite{mkn,SME,michela} as well as its  various extended versions \cite{BSM,majorana,thdm,ESMCO,elptdm,phonglongvan,SMS,dssm,munusm,lr,singlet,mssm1,twostep,1101.4665}.
For the SM, although the EWPT strength is larger than unity at the electroweak scale,  the mass of the Higgs boson must be less than $125$ GeV \cite{mkn,SME,michela}; so the EWBG requires new physics beyond the SM at the weak scale \cite{BSM}.

Many extensions such as the two-Higgs-boublet model, the reduced minimal 3-3-1 model, the economical 3-3-1 model or
 the Minimal Supersymmetric Standard Model, have a strongly first-order EWPT and the new sources of CP violation,
 which are necessary to account for the BAU; triggers for the first-order EWPT in these models are heavy bosons or Dark Matter candidates \cite{majorana,thdm,ESMCO,elptdm,phonglongvan,singlet,mssm1,twostep, chiang3}. However,  most research of the EWPT
  are the Landau gauge. Recently gauge invariant also made important contributions in the EWPT as researching in Refs.\cite{1101.4665,Arefe}.

The quantity of sphaleron rate
admitting to  B violation rate, has been calculated in the SM in Refs.\cite{mkn,SME,michela} and in the
 reduced minimal 3-3-1 model in Ref. \cite{phonglongvan}. In addition, by using nonperturbative lattice simulations,
 a powerful framework and set of analytic and numerical tools have been developed in Refs. \cite{SME,michela}.

The Zee-Babu (ZB) model is one of  the simplest extensions of the SM which has some interesting features \cite{zeebabu}.
 Due to its simplicity, in this work,  we have considered the EWPT and sphaleron rate in the  ZB model.

In the ZB model,  two extra  charged scalars $h^{\pm}$ and $k^{\pm\pm}$ are added  to the Higgs potential.
The kind of new scalars can play an important role in the early Universe.  As shown in \cite{zeebabu,bzee}, they can also be a reason for tiny mass of neutrinos
 through two loop or three loop corrections. One important property of these particles which will be shown in this paper,
  is that they can be triggers for the first-order phase transition.

In order to drive a gauge dependent effective potential at one-loop level,  in this paper we will use a direct method which is different
 from those used in Refs. \cite{1101.4665, Arefe}. This paper  is organized as follows. In Sec. \ref{sec2} we give a short review of
 the ZB model and we drive an effective potential which has a contribution from heavy scalars and the $\xi$ gauge at one-loop level.
 In Sec. \ref{sec3}, we find the  mass range  of charged scalar particles by a first-order phase transition condition. Finally, Sec. \ref{sec5}
  is devoted to constraints on the mass of the charged Higgs boson. In Sec. \ref{sec6} we summarize and describe outlooks.

\section{EFFECTIVE POTENTIAL IN THE ZEE-BABU MODEL}\label{sec2}

In the ZB model, by adding two charged scalar fields $h^{\pm}$ and $k^{\pm\pm}$ \cite{zeebabu}, the Lagrangian becomes
\bea\label{lzb}
\mathcal{L}  &=& L_{SM}+f_{ab}\overline{\psi_{aL}^c}\psi_{bL}h^{+} +h_{ab}^{'}\overline{l_{aR}^c}l_{bR}k^{++} + V(\phi,h,k)
\crn
 & &+ (D_{\mu}h^+)^{\dagger}(D^{\mu}h^{+}) +
(D_{\mu}k^{++})^{\dagger}(D^{\mu}k^{++})+H.c
\label{1}
\eea
In the model, the Higgs potential contains more four couplings between $h^{\pm}$ or $k^{\pm\pm}$ and neutral Higgs  boson \cite{zeebabu}:
\bea
V(\phi,h,k)&=&\mu^2\phi^{\dagger}\phi + u^2_1{\vert{h}\vert}^2 + u_2^2{\vert{k}\vert}^2 +\lambda (\phi^{\dagger}\phi)^2 +
\lambda_h{ \vert{h}\vert}^4 + \lambda_k{ \vert{k}\vert}^4
\crn
& &  +\lambda_{hk}{ \vert{h}\vert}^2{ \vert{k}\vert}^2 + 2p^2{\vert{h}\vert}^{2} \phi^{\dagger}\phi
 + 2q^2{\vert{k}\vert}^{2} \phi^{\dagger}\phi + (\mu_{hk} h^{2} k^{++} + H.c)\, ,
 \label{2}
\eea
where
\be
\phi=\left(\begin{array}{c}
\rho^+ \\
\rho^0
\end{array}\right)
\label{3}
\ee
and $\rho^0$  has a vacuum expectation value (VEV)
\be
\rho^{0}=\frac{1}{\sqrt{2}}\left( v_0+\sigma+i\zeta\right)\, .\label{4}
\ee

The masses of $h^\pm$ and $k^{{\pm\pm}}$ are given by
\bea\label{mass}
m^2_{h^\pm} &= &p^2v^2_0+u^2_1,\crn
m^2_{{k^{\pm\pm}}}&=&q^2v^2_0+u^2_2.
\eea

Diagonalizing matrices in the kinetic components of the Higgs potential and retaining Goldstone bosons, we obtain
\begin{gather}\label{eq:SMfieldDepmasses}
\begin{aligned}
m_H^2(v_0)&=-\mu^2+3\lambda v^2_0\,,\\
m_G^2(v_0)&=-\mu^2+\lambda v^2_0\,,\\
\end{aligned}
\begin{aligned}
m_Z^2(v_0)&=\textstyle\frac{1}{4}(g^2+g'^2) v^2_0=a^2v^2_0\,,\\
m_W^2(v_0)&=\textstyle\frac{1}{4}g^2 v^2_0=b^2v^2_0\,.\\
\end{aligned}
\end{gather}

\subsection{EFFECTIVE POTENTIAL WITH LANDAU GAUGE}

From Eq. (\ref{lzb}), ignoring Goldstone bosons, we  obtain an effective potential with contributions of $h^{\pm}$ and $k^{{\pm\pm}}$ in the Landau gauge:
\bea\label{pf}
 V _{eff}(v)&= &V _{0}(v)+\frac{3}{64\pi^2}\left( m _{Z}^{4}(v)ln\frac{m _{Z}^{2}(v)}{Q^2}
+2m _{W}^{4}(v)ln\dfrac{m _{W}^{2}(v)}{Q^2} - 4m _{t}^{4}(v)ln\dfrac{m _{t}^{2}(v)}{Q^2}\right)\crn
 &+& \frac{1}{64\pi^2}\left(2m_{h^\pm}^{4}(v)ln\dfrac{m_{h^{\pm}}^{2}(v)}{Q^2} + 2m_{k^{\pm\pm}}^{4}(v){ln}\dfrac{m_{k^{\pm\pm}}^{2}(v)}{Q^2}+ m _ {H}^{4}(v)ln\dfrac{m_{H}^{2}(v)}{Q^2} \right)\crn
&+& \frac{3 T^4}{4\pi^2} \left\{F_{-} (\frac{m_Z(v)}{T})+F_{-}(\frac{m_W(v)}{T}) + 4F_{+}(\frac{m_t(v)}{T})\right\}\crn
&+& \frac{T^4}{4\pi^2}\left\{2{F}_{-}(\frac{m_{h^{\pm}}(v)}{T})+2{F}_{-}(\frac{m_{k^{\pm\pm}}(v)}{T}) + {F}_{-}(\frac{{m}_{H}(v)}{T})\right\}\, ,
\eea
where $v_\rho$ is a variable changing with temperature, and at  $T=0$,  $v_\rho \equiv v_0=246$ GeV. Here
\bea
F_{\pm}\left(\dfrac{m_\phi}{T}\right) & = &\int_{0}^{\dfrac{m_\phi}{T}}\alpha J_{\mp}^{1} (\alpha,0)d\alpha,\crn
 J_{\mp}^{1} (\alpha,0)&= & 2\int_{\alpha} ^{\infty}\dfrac{(x^2-\alpha^ 2)^{\frac{1}{2}}}{e^x\mp 1}dx.\nn
 \eea

\subsection{EFFECTIVE POTENTIAL WITH $\xi$ GAUGE}

 It is known that in high levels, the contribution of Goldstone boson cannot be ignored. Therefore, we must consider an effective
  potential in arbitrary $\xi$ gauge given by
\bea\label{Veff0}
\mathcal{V}_1^{T=0}(v)&=&\frac{1}{4(4\pi)^2}(m_H^2)^2\big[\ln(\textstyle\frac{m_H^2}{Q^2})-\frac{3}{2}\displaystyle\big]+
\frac{1}{4(4\pi)^2}(m_{h^{\pm}}^2)^2\big[\ln(\textstyle\frac{m_{h^\pm}^2}{Q^2})-\frac{3}{2}\displaystyle\big]\crn
&&
+\frac{1}{4(4\pi)^2}(m_{k^{\pm\pm}}^2)^2\big[\ln(\textstyle\frac{m_{{k^{\pm\pm}}}^2}{Q^2})-\frac{3}{2}\displaystyle\big]+
\frac{2\times1}{4(4\pi)^2}(m_G^2+\xi m_W^2)^2\big[\ln(\textstyle\frac{m_G^2+\xi m_W^2}{Q^2})-\frac{3}{2}\displaystyle\big]\crn
&&
+\frac{1}{4(4\pi)^2}(m_G^2+\xi m_Z^2)^2\big[\ln(\textstyle\frac{m_G^2+\xi m_Z^2}{Q^2})-\frac{3}{2}\displaystyle\big]
+\frac{2\times3}{4(4\pi)^2}(m_W^2)^2\big[\ln(\textstyle\frac{m_W^2}{Q^2})-\frac{5}{6}\displaystyle\big]\crn
&&
+\frac{3}{4(4\pi)^2}(m_Z^2)^2\big[\ln(\textstyle\frac{m_Z^2}{Q^2})-\frac{5}{6}\displaystyle\big]
-\frac{2\times1}{4(4\pi)^2}(\xi m_W^2)^2\big[\ln(\textstyle\frac{\xi m_W^2}{Q^2})-\frac{3}{2}\displaystyle\big]\crn
&&
-\frac{1}{4(4\pi)^2}(\xi m_Z^2)^2\big[\ln(\textstyle\frac{\xi m_Z^2}{Q^2})-\frac{3}{2}\displaystyle\big]-\text{``free''},
\eea
and
\bea\label{Veff}
\mathcal{V}^{T\neq0}_1(v,T)&=&\frac{T^4}{2\pi^2}\bigg[J_B\Big(\frac{m_H^2}{T^2}\Big)+J_B\Big(\frac{m_{h^{\pm}}^2}{T^2}\Big)
+2J_B\Big(\frac{m_{k^{\pm\pm}}^2}{T^2}\Big)\bigg]\crn
&&
+\frac{T^4}{2\pi^2}\bigg[2\!\times\!J_B\Big(\frac{m_G^2+\xi m_W^2}{T^2}\Big)+J_B\Big(\frac{m_G^2+\xi m_Z^2}{T^2}\Big)\bigg]\crn
&&
+\frac{3T^4}{2\pi^2}\bigg[2\!\times\!J_B\Big(\frac{m_W^2}{T^2}\Big)+J_B\Big(\frac{m_Z^2}{T^4}\Big)+
J_B\Big(\frac{m_\gamma^2}{T^4}\Big)\bigg]\crn
&&
-\frac{T^4}{2\pi^2}\bigg[2\!\times\!J_B\Big(\frac{\xi m_W^2}{T^2}\Big)+J_B\Big(\frac{\xi m_Z^2}{T^2}\Big)+J_B\Big(\frac{\xi m_\gamma^2}{T^2}\Big)\bigg]-\text{``free''},
\eea
where ``free'' represents a free-field subtraction.

\section{ELECTROWEAK PHASE TRANSITION IN THE ZEE-BABU MODEL}\label{sec3}
\subsection{EWPT in Landau gauge}
Ignoring $u_1$ and $u_2$ in Eq.(\ref{mass}) (i.e., $u_1$ and $u_2$ are assumed to be very small) and neglecting contributions of
 Goldstone bosons, we can write the high-temperature expansion of the potential in  Eq.(\ref{pf}) as a quartic
expression in $v$:
\be\label{EP-v-1}
V_{eff}(v)=D(T^2-T^2_0)v^2-ET|v|^3+\frac{\lambda_T}{4}v^4,
\ee
in which
\bea
D &=&\frac{1}{24 {v_0}^2} \left[
	6 m_W^2(v_0) +3m_{Z_1}^2(v_0)
	+m_{H}^2(v_0)+2m_{h^\pm}^2(v_0) +2m_{k^{\pm\pm}}^2(v_0) +6 m_t^2 (v_0)
 	\right],\crn
T_0^2&=&\frac{1}{D}\left\{\frac{m_{H}^2(v_0)}{4}
	-\frac{1}{32\pi^2v_0^2}\left(6m_W^4(v_0)+3m_{Z_1}^4(v_0)+m_{H}^4(v_0)
\right.\right.\crn
	&&\left.\left.\qquad \qquad \qquad \qquad+2m_{h^\pm}^4(v_0) +2m_{k^{\pm\pm}}^4(v_0) -12 m_t^4 (v_0)
\right)\right\},\crn
E &=& \frac{1}{12 \pi v_0^3} \left(
		6 m_W^3(v_0) +3m_{Z_1}^3(v_0)
		+m_{H}^3(v_0) +2m_{h^\pm}^3(v_0) +2m_{k^{\pm\pm}}^3(v_0)
	\right),\label{D-T-E-lambda}\\
\lambda_T &=&
 \frac{m_{H}^2(v_0)}{2 v_0^2}\left\{
 	1- \frac{1}{8\pi^2 v_0^2 (m_{H}^2(v_0))}\left[
 		6 m_W^4(v_0) \ln \frac{m_W^2(v_0)}{a_b T^2} 		
 	\right.\right.\crn
	&&\qquad \left.\left.
+3m_{Z_1}^4(v_0) \ln\frac{m_{Z_1}^2(v_0)}{a_b T^2}+m_{H^0}^4(v_0) \ln \frac{m_{H^0}^2(v_0)}{a_b T^2}		
		\right.\right.\crn
&&\qquad \left.\left.
		+2m_{k^{\pm\pm}}^4(v_0)\ln\frac{m_{k^{\pm\pm}}^2(v_0)}{a_bT^2}+2m_{h^\pm}^4(v_0)\ln\frac{m_{h^\pm}^2(v_0)}{a_bT^2}-12 m_t^4(v_0)\ln\frac{m_t^2(v_0)}{a_F T^2}
	\right] \right\},\nn \eea
where $v_0$ is the value where the zero-temperature effective potential $V^{0}_{eff}(v)$ gets the minimum. Here, we acquire $V^{0}_{eff}$ from $V_{eff}$ in Eq.\eqref{EP-v-1} by neglecting all terms in the form $F_{\mp}\left(\frac{m}{T}\right)$.

The minimum conditions for $V^{0}_{eff}(v)$ are
\be\label{MinCondition-v}
V^{0}_{eff}(v_0)=0,\hs
\frac{\partial V^{0}_{eff}(v)}{\partial v}\Big|_{v=v_0}=0,\hs
 \frac{\partial^2 V^{0}_{eff}(v)}{\partial v^2}\Big|_{v=v_0}=
 \left[m^2_{H}(v)\right] \Big|_{v=v_0}=125^2\text{ GeV}^2.
\ee
We also have the minima of the effective potential in Eq.\eqref{EP-v-1}
\be \label{v-0}
v=0, \quad v \equiv v_c=\frac{2ET_c}{\lambda_{T_c}},
\ee
where $v_c$ is the critical VEV of  $\phi$ at the broken state, and $T_c$ is the critical temperature of phase transition given by
\be \label{Tc}
T_c=\frac{T_0}{\sqrt{1-E^2/D\lambda_{T_c}}}.
\ee

\begin{figure}[!ht]
\centering
\includegraphics[height=9cm, width=16cm]{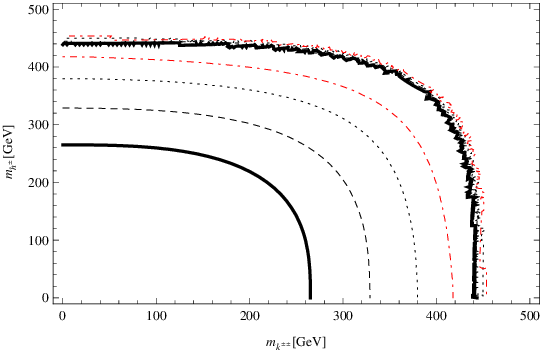}
\caption{When the solid contour of $S=2E/\lambda_{T_c}=1$, the
dashed contour: $2E/\lambda_{T_c}=1.5$, the dotted contour: $2E/\lambda_{T_c}=2$,
the dotted-dashed contour: $2E/\lambda_{T_c}=2.4$, even and nosmooth contours: $S\longrightarrow \infty$.}\label{fig:01}
\end{figure}

Now let us investigate the phase transition strength
\be \label{S}
S=\frac{v_{c}}{T_c}=\frac{2E}{\lambda_{T_c}}
\ee
of this EWPT. In the limit $E \rightarrow 0$, the transition strength tends to zero ($S\rightarrow 0$) and the phase transition is a second-order one.
To have a first-order phase transition, we require that the strength is larger or equal to the unit ($S \geq 1$).  In Fig. \ref{fig:01}, we have  plotted the transition strength $S$ as a function of the new charged scalars: $m_{h^{\pm}}$ and $m_{k^{\pm\pm}}$.

According to Ref. \cite{5percent}, the accuracy of a high-temperature expansion for the effective potential such as that in Eq. \eqref{EP-v-1}
 will be better than $5\%$ if $\frac{m_{boson}}{T}< 2.2$, where $m_{boson}$ is the relevant boson mass. Therefore, as shown in
  Fig. \ref{fig:01}, for $m_{h^{\pm}}$ and $m_{k^{\pm\pm}}$
   being  in the  $0-350\, \mathrm{GeV}$ range, respectively,
  the transition strength is in the range $1 \leq S <2.4$.

We see that the contribution of $h^{\pm}$ and $k^{\pm\pm}$ are the same. The larger mass of $h^{\pm}$ and $k^{\pm\pm}$,
the larger cubic term ($E$) in the effective potential but the strength of phase transition cannot be strong. Because the value of $\lambda$
 also increases, so there is a tension between $E$ and $\lambda$ to make the first order phase transition. In addition when the masses
  of charged Higgs bosons are too large, $T_0, \lambda$ will be unknown or $S\longrightarrow \infty$.

\subsection{EWPT in $\xi$ gauge}

The high-temperature expansions of the potential in Eq.(\ref{Veff0}) and in Eq.(\ref{Veff}) can be rewritten in   a like-quartic expression in $v$
\bea
\mathcal{V}&=&\left(\mathcal{D}_1+\mathcal{D}_2+\mathcal{D}_3+\mathcal{D}_4+\mathcal{B}_2\right)v^2+\mathcal{B}_1v^3
+\Lambda v^4+f(T,u_1,u_2,\mu,\xi), \label{vxi}
\eea
where
\bea\label{ct0}
f(T,u_1,u_2,\mu,\xi, v)=\mathcal{C}_1+\mathcal{C}_2,\eea
and
\bea\label{ct1}
\mathcal{D}_1&=&\frac{T^2}{24 v^2_0} \left(3m^2_Z(v_0)+6 m^2_W(v_0)+6 m^2_t(v_0)+2(m^2_{h^\pm}(v_0)-u_1^2)
+2(m^2_{k^{\pm\pm}}(v_0)-u_2^2)+6\lambda v^2_0\right)\, ,\crn
\mathcal{D}_2&=&\frac{1}{32 v^2_0\pi^2}\left\{3m^4_Z(v_0)+6m^4_W(v_0)-12m^4_t(v_0)+2(m^2_{h^\pm}(v_0)-u_1^2)^2
-8\pi^2 v^2_0 m^2_{H_0}\right.\crn
&&\quad \left.
+2(m^2_{k^{\pm\pm}}(v_0)-u_2^2)^2+12v_0^4\lambda^2+2m^2_Z(v_0)v^2_0\lambda\xi +4m^2_W(v_0)v^2_0\lambda\xi\right\},\crn
\mathcal{D}_3&=&\frac{1}{32\pi^2}\left\{2p^2 u_1^2 \ln\left[\frac{a_b T^2}{p^2 v_0^2+u_1^2}\right]+2 q^2 u_2^2 \ln\left[\frac{a_b T^2}{q^2 v_0^2+u_2^2}\right]\right.\crn
&&\quad \left.
-3\lambda\mu^2 \ln\left[\frac{a_b T^2}{3v_0^2\lambda-\mu^2}\right]-\lambda\mu^2 \ln\left[\frac{a_b T^2}{v_0^2 \left(\lambda
+a^2\xi \right)-\mu^2}\right]\right.\crn
&&\quad \left.
-2\lambda\mu^2 \ln\left[\frac{a_bT^2}{v_0^2 \left(\lambda +b^2\xi \right)-\mu^2}\right]-a^2 \xi\mu^2 \ln\left[\frac{a_b T^2}{v_0^2 \left(\lambda +a^2 \xi \right)-\mu^2}\right]\right.\crn
&&\quad \left.
-2b^2 \xi\mu^2\ln\left[\frac{a_bT^2}{v_0^2\left(\lambda+b^2 \xi \right)-\mu^2}\right]\right\}\,,\crn
\mathcal{D}_4&=&\frac{1}{32\pi^2}\left(2p^2u_1^2+2q^2 u_2^2-6\lambda\mu^2-a^2\xi\mu^2-2b^2\xi\mu^2\right)\, ,\nn
\eea
\bea
\Lambda&=&\frac{1}{64\pi^2}\left\{2p^4 \ln\left[\frac{a_bT^2}{u_1^2+p^2v_0^2}\right]+2q^4 \ln\left[\frac{a_bT^2}{u_2^2+q^2v_0^2}\right]+3a^4\ln\left[\frac{a_bT^2}{a^2v_0^2}\right]+6b^4\ln\left[\frac{a_bT^2}{b^2v_0^2}\right]\right.\crn
&&\quad \left.
-12k^4 \ln\left[\frac{a_F T^2}{k^2v_0^2}\right]+9\lambda^2\ln\left[\frac{a_bT^2}{3\lambda v_0^2-\mu^2}\right]
+8\pi^2 \frac{m^2_{H_0}}{v_0^2}\right.\crn
&&\quad \left.
-a^4\xi^2\ln\left[\frac{a_bT^2}{a^2\xi v_0^2}\right]-2b^4\xi^2\ln\left[\frac{a_bT^2}{b^2\xi v_0^2}\right]\right.\crn
&&\quad \left.
+a^4\xi^2\ln\left[\frac{a_b T^2}{v_0^2 \left(\lambda +a^2 \xi \right)-\mu^2}\right]+2b^4\xi^2\ln\left[\frac{a_b T^2}{v_0^2\left(\lambda +b^2\xi \right)-\mu^2}\right]\right.\crn
&&\quad \left.
+2a^2\lambda\xi\ln\left[\frac{a_b T^2}{v_0^2\left(\lambda+a^2\xi \right)-\mu^2}\right]+4b^2\lambda\xi\ln\left[\frac{a_b T^2}{v_0^2 \left(\lambda+b^2\xi \right)-\mu^2}\right]\right.\crn
&&\quad \left.
+\lambda^2\ln\left[\frac{a_b T^2}{v_0^2\left(\lambda+a^2\xi \right)-\mu^2}\right]+2\lambda^2\ln\left[\frac{a_b T^2}{v_0^2\left(\lambda
+b^2\xi \right)-\mu^2}\right]\right\}\, ,\nn
\eea
\bea
\mathcal{B}_1&=&\frac{T}{12 \pi v^3_0}\left(-3m^3_Z(v_0)-6m^3_W(v_0)+m^3_Z(v_0)\xi^{3/2}+2m^3_W(v_0)\xi^{3/2}
\right)\, ,\crn
\mathcal{B}_2&=&T\left(-\frac{p^2 \sqrt{u_1^2+p^2 v^2}}{6 \pi }-\frac{q^2\sqrt{u_2^2+q^2 v^2}}{6 \pi }-
\frac{\lambda\sqrt{3\lambda v^2-\mu ^2}}{4\pi }-\frac{\lambda\sqrt{\lambda v^2+a^2 \xi v^2-\mu^2}}{12\pi}\right.\crn
&&\quad \left.
-\frac{a^2\xi\sqrt{\lambda v^2+a^2\xi v^2-\mu^2}}{12\pi }-\frac{\lambda\sqrt{\lambda v^2+b^2\xi v^2-\mu^2}}{6\pi }-
\frac{b^2\xi\sqrt{\lambda v^2+b^2 \xi v^2-\mu^2}}{6\pi }\right)\, ,
\eea
\bea\label{ct2}
\mathcal{C}_1&=&-\frac{Tu_1^2\sqrt{u_1^2+p^2 v^2}}{6\pi }-\frac{T u_2^2\sqrt{u_2^2+q^2 v^2}}{6 \pi }-\frac{T^2\mu^2}{6}
+\frac{3\mu^4}{32\pi^2}\crn
&+&\frac{T \mu^2\sqrt{3\lambda v^2-\mu^2}}{12\pi}+\frac{T \mu^2\sqrt{\lambda v^2+a^2\xi v^2-\mu^2}}{12\pi}
+\frac{T\mu^2 \sqrt{\lambda v^2+b^2 \xi v^2-\mu^2}}{6\pi }\crn
&+&\frac{u_1^4\ln\left[\frac{a_b T^2}{v_0^2}\right]}{32\pi^2}+\frac{u_2^4 \ln\left[\frac{a_b T^2}{ v_0^2}\right]}{32\pi^2}
+3\frac{\mu^4 \ln\left[\frac{a_bT^2}{v_0^2}\right]}{64\pi^2}\, ,\crn
\mathcal{C}_2&=&\frac{T^2u_1^2}{12}+\frac{3u_1^4}{64\pi^2}+\frac{T^2 u_2^2}{12}+\frac{3u_2^4}{64\pi ^2}+\delta\Omega\, ,
\nn
\eea
\bea
\delta\Omega&=&-\frac{1}{128\pi^2}\left(-4p^2 u_1^2 v_0^2-4q^2u_2^2 v_0^2+3a^4v_0^4+6b^4v_0^4-12k^4v_0^4
+2p^4v_0^4+2q^4 v_0^4\right.\crn
&&\quad \left.
+12v_0^4\lambda^2+2a^2v_0^4\lambda\xi+4b^2v_0^4\lambda\xi+12v_0^2\lambda\mu^2+2a^2v_0^2\xi\mu^2+4b^2v_0^2\xi\mu^2\right.\crn
&&\quad \left.
+4 u_1^4\ln\left[\frac{u_1^2+p^2v_0^2}{v_0^2}\right]+4u_2^4 \ln\left[\frac{u_2^2+q^2v_0^2}{v_0^2}\right]+
2\mu^4 \ln\left[\frac{3 v_0^2\lambda -\mu^2}{v_0^2}\right]\right.\crn
&&\quad \left.
+2\mu^4 \ln\left[\frac{v_0^2\lambda+a^2v_0^2\xi -\mu^2}{v_0^2}\right]+4\mu^4\ln\left[\frac{v_0^2\lambda+b^2 v_0^2 \xi -\mu^2}{v_0^2}\right]-16\pi^2v_0^2m^2_{H_0}\right).
\eea

Expanding functions $J_B\Big(\frac{m_G^2+\xi m_W^2}{T^2}\Big)$ and $J_B\Big(\frac{m_G^2+\xi m_Z^2}{T^2}\Big)$ in Eq. (\ref{Veff}),
we will obtain the  term of  mixing between $\xi$ and $v$ in $\mathcal{B}_1$ and $\mathcal{B}_2$. Therefore $J_B\Big(\frac{m_G^2
+\xi m_W^2}{T^2}\Big)$ and $J_B\Big(\frac{m_G^2+\xi m_Z^2}{T^2}\Big)$ or $\mathcal{B}_1$ and $\mathcal{B}_2$ contain
a part of daisy diagram contributions mentioned in Ref. \cite{1101.4665}. The other part of ring-loop distribution comes to damping effect.
The damping effect is in the thermal self-energy term $(\Sigma_{ij}(T)\phi_i\phi_j$ and $\Pi^{ab}(T)A^a_0A^b_0$, i.e., $V^B_{ring}$ in
 Ref. \cite{1101.4665}).

On the other hand, we see that the ring loop distribution still is very small, it was approximated $g^2T^2/m^2$ ($g$ is the
coupling constant of $SU(2)$, $m$ is mass of boson), $m\sim 100$ GeV, $g\sim 10^{-1}$ so $g^2/m^2 \sim 10^{-5}$. If we add this
 distribution to the effective potential, the $\mathcal{D}_1$ term will give a small change  only. Therefore, this distribution
 does not change the strength of EWPT or, in other words, it  is not the origin of EWPT.

The potential in  Eq.(\ref{vxi}) is not  a quartic expression because $\mathcal{B}_2, \mathcal{D}_3, \mathcal{D}_4$ and $f(T,u_1,u_2,\mu,\xi, v)$
 depend on $v$, $\xi$ and $T$. It has seven variables such as $u_1, u_2, p, q, \mu, \lambda$ and $ \xi$. Therefore, the shape of potential is
  distorted by $u_1, u_2, p,q, \xi$ but not so much.  If Goldstone bosons are neglected and the gauge parameter is  vanished ($\xi=0$), it will be reduced to
  Eq.(\ref{EP-v-1}) in the Landau gauge.

The minimum conditions for Eq.(\ref{vxi}) are still like Eq.(\ref{MinCondition-v}) but for this case, it holds: $m^2_{H_0}=-\mu^2+3\lambda v^2_0=125^2 \text{ GeV}$.

There are many variables in our problem and some of them, for example, $u_1, u_2, p, q$ and $ \mu$ play the same role. They are components
 in the mass of particles.

It is emphasized that $\xi$ and $ \lambda$ are two important variables and have different roles. Therefore, in order to reduce number of variables, we have to approximate values of variables, but must not lose the generality of the problem and simplify $\mathcal{B}_2, \mathcal{D}_3, \mathcal{D}_4, f(T,u_1,u_2,\mu,\xi, v)$ in the next section.

\subsection{The case of small contribution of Goldstone boson}

When the mass of Goldstone boson is small, i.e., $\mu^2 \approx \lambda v_0^2$ and taking into account $m_{H_0}= 125$ GeV,
    we obtain $\lambda=0.1297$. Note that this is a consequence of the  above argument, in which the values $u_1$ and $ u_2$ are ignored
 because their existence deforms the potential.

In this sub-section, proving the gauge independent effective potential, we conduct a method
yielding  an effective potential as
 a quartic expression in $v$ through three steps.

The first approximate step is as follows: when $ \mu^2 \approx \lambda v_0^2$, the term $6\lambda v^2_0$ in $\mathcal{D}_1$
 can be simplified with $-\frac{T^2\mu^2}{6}$ in $\mathcal{C}_1$. All terms in $\mathcal{D}_4$ will be destroyed so that $\mathcal{D}_1$
  and $\mathcal{D}_2$ can be rewritten as
\bea\label{ct3}
\mathcal{D}_1&=&\frac{T^2}{24 v^2_0} \left(3m^2_Z(v_0)+6 m^2_W(v_0)+6 m^2_t(v_0)+2m^2_{h^\pm}(v_0)+2m^2_{k^{\pm\pm}}(v_0)
+2\lambda v^2_0\right) ,\crn
\mathcal{D}_2&=&\frac{1}{32 v^2_0\pi^2}\left\{3m^4_Z(v_0)+6m^4_W(v_0)-12m^4_t(v_0)+2(m^2_{h^\pm}(v_0)-u_1^2)^2-
8\pi^2 v^2_0 m^2_{H_0}\right.\crn
&&\quad \left.
+2(m^2_{k^{\pm\pm}}(v_0)-u_2^2)^2+6v_0^4\lambda^2+m^2_Z(v_0)v^2_0\lambda\xi +2m^2_W(v_0)v^2_0\lambda\xi\right\} .
\eea

In the second approximate step,  we neglect $u_1, u_2$, and obtain
\bea\label{ct4}
\mathcal{D}_3&=&\frac{1}{32\pi^2}\left\{-3\lambda\mu^2 \ln\left[\frac{a_b T^2}{2v_0^2\lambda}\right]-\lambda\mu^2
\ln\left[\frac{a_b T^2}{v_0^2 \left(a^2\xi \right)}\right]\right.\crn
&&\quad \left.
-2\lambda\mu^2 \ln\left[\frac{a_bT^2}{v_0^2 \left(b^2\xi \right)}\right]-a^2 \xi\mu^2 \ln\left[\frac{a_b T^2}{v_0^2 \left(a^2 \xi \right)}\right]\right.\crn
&&\quad \left.
-2b^2 \xi\mu^2\ln\left[\frac{a_bT^2}{v_0^2\left(b^2 \xi \right)}\right]\right\}\,,\crn
\Lambda&=&\frac{1}{64\pi^2}\left\{2p^4 \ln\left[\frac{a_bT^2}{p^2v_0^2}\right]+2q^4 \ln\left[\frac{a_bT^2}{q^2v_0^2}\right]+3a^4\ln\left[\frac{a_bT^2}{a^2v_0^2}\right]+6b^4\ln\left[\frac{a_bT^2}{b^2v_0^2}\right]\right.\crn
&&\quad \left.
-12k^4 \ln\left[\frac{a_F T^2}{k^2v_0^2}\right]+9\lambda^2\ln\left[\frac{a_bT^2}{3\lambda v_0^2-\mu^2}\right]+8\pi^2 \frac{m^2_{H_0}}{v_0^2}\right.\crn
&&\quad \left.
+2a^2\lambda\xi\ln\left[\frac{a_b T^2}{v_0^2\left(a^2\xi \right)}\right]+4b^2\lambda\xi\ln\left[\frac{a_b T^2}{v_0^2 \left(b^2\xi \right)}\right]\right.\crn
&&\quad \left.
+\lambda^2\ln\left[\frac{a_b T^2}{v_0^2\left(a^2\xi \right)}\right]+2\lambda^2\ln\left[\frac{a_b T^2}{v_0^2\left(b^2\xi \right)}\right]\right\}\, ,\nn
\eea
\bea
\mathcal{B}_1&=&\frac{T}{12 \pi v^3_0}\left(-3m^3_Z(v_0)-6m^3_W(v_0)+m^3_Z(v_0)\xi^{3/2}+2m^3_W(v_0)\xi^{3/2}
\right)\, ,\crn
\mathcal{B}_2&=&T\left(-\frac{p^2 \sqrt{p^2 v^2}}{6 \pi }-\frac{q^2\sqrt{q^2 v^2}}{6 \pi }-\frac{\lambda\sqrt{3\lambda v^2-\mu ^2}}{4\pi }-\frac{\lambda\sqrt{\lambda v^2+a^2 \xi v^2-\mu^2}}{12\pi}\right.\crn
&& \left.
-\frac{a^2\xi\sqrt{\lambda v^2+a^2\xi v^2-\mu^2}}{12\pi }-\frac{\lambda\sqrt{\lambda v^2+b^2\xi v^2-\mu^2}}{6\pi }-\frac{b^2\xi\sqrt{\lambda v^2+b^2 \xi v^2-\mu^2}}{6\pi }\right)\, ,\nn
\eea
\bea
\mathcal{C}_1&=&-\frac{T^2\mu^2}{6}+\frac{3\mu^4}{32\pi^2}\crn
	&+&\frac{T \mu^2\sqrt{3\lambda v^2-\mu^2}}{12\pi}+\frac{T \mu^2\sqrt{\lambda v^2+a^2\xi v^2-\mu^2}}{12\pi}+\frac{T\mu^2 \sqrt{\lambda v^2+b^2 \xi v^2-\mu^2}}{6\pi }\crn
	&+&3\frac{\mu^4\ln\left[\frac{a_bT^2}{v_0^2}\right]}{64\pi^2}\, ,\crn
	\delta\Omega&=&-\frac{1}{128\pi^2}\left(3a^4v_0^4+6b^4v_0^4-12k^4v_0^4+2p^4v_0^4+2q^4 v_0^4\right.\crn
	&&\quad \left.
	+12v_0^4\lambda^2+2a^2v_0^4\lambda\xi+4b^2v_0^4\lambda\xi+12v_0^2\lambda\mu^2+2a^2v_0^2\xi\mu^2+4b^2v_0^2\xi\mu^2\right.\crn
	&&\quad \left.
	+2\mu^4\ln\left[2\right]+2\mu^4\ln\left[a^2\xi\right]+4\mu^4\ln\left[b^2\xi\right]-16\pi^2v_0^2m^2_{H_0}\right) .
\eea

In the third approximate step:  replacing $\mu^2=\lambda v^2_0$ and in the square root term of $\mathcal{B}_2$ and $\mathcal{C}_1$,
we can approximate $\mu^2 \sim \lambda v^2$. Therefore, all terms in $\mathcal{C}_1$ are destroyed with $\delta\Omega$ (except the last one, $\fr{3}{64\pi^2}\mu^4\ln\left[\frac{a_bT^2}{v_0^2}\right]$) and $\mathcal{D}_3$ and $\mathcal{B}_2$, will finally be simplified
with $\Lambda$ and $\mathcal{B}_1$, respectively.

The value $\fr{3}{64\pi^2}\mu^4\ln\left[\frac{a_bT^2}{v_0^2}\right]$ depends on $T$, and it pushes the effective potential to right,
or it distorts the quadratic potential and this shows the effect of $\xi$ as seen in Ref.\cite{1101.4665}. Therefore the mentioned term
in $\mathcal{C}_1$ can be neglected.

\begin{figure}[!ht]
	\centering
	\includegraphics[height=9cm, width=16cm]{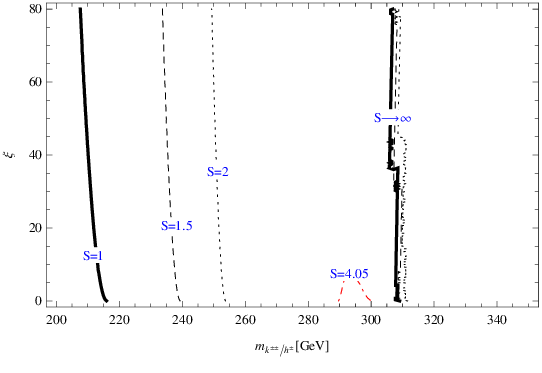}
	\caption{The strength of EWPT with $\lambda=0.1297$ and $\mu^2 \sim \lambda v^2_0$ }\label{fig:07}
\end{figure}

Finally, we obtain the strength of EWPT as shown in Fig.\ref{fig:07}. The maximum of the strength is about 4.05.

In fact, the mass of Goldstone boson is much smaller than that of the $W^{\pm}$ boson  or the $Z$ boson so the contribution of Goldstone boson
must be very small in the effective potential. Hence, the lines in Fig.\ref{fig:07} are almost vertical or almost parallel to the axis $\xi$.
These results match those of Ref.\cite{Arefe}. This shows that \emph{the strength of EWPT is gauge independent}.

In addition, the new particles have large masses, so they
provide valuable contributions to the EWPT in the Landau gauge or in  an arbitrary gauge.
The charge of these particles increases their contributions. In particular, $k^{\pm\pm}$ is the doubly charged scalar, so its
coefficients in the effective potential are also greater than two times the coefficient of $W^{\pm}$ boson
 (because of the fact that  the  doubly charged particle
 ($k^{\pm\pm}$)  always appears  in pairs with the singly charged one ($h^{\pm}$), and by our  approximation,    with the same masses: $m_{k^{\pm\pm}}=m_{h^{\pm}}$).
	
Furthermore, we find that models
having  doubly charged particles, provide  a very strong first-order EWPT, such as the
 Georgi–Machacek model \cite{chiang3} and they are being tested by LHC \cite{chiang1, chiang2}.

 According to Nielsen's identity, in $\hbar$ expansion,   the one-loop effective potential  is gauge
 independent at each order by $\hbar$ \cite{1101.4665},  but the  general potential still is gauge dependent.
  However this dependence  is not important as in our above  analysis.

\section{Constraints on  coupling constants in the Higgs potential}\label{sec5}

In order to have the first order phase transition, the masses of the new  charged scalars $m_{h^{\pm}}$ and $m_{k^{\pm\pm}}$ must be smaller than $350$ GeV. Therefore, we obtain
\be
p^2v^2_0<(300 \text{ GeV})^2, \label{re1}
\ee
and
\be
q^2v^2_0<(300 \text{ GeV})^2\, .\label{re2}
\ee

From the above equations, we obtain the following limits:  $0<p<1.22$ and $0<q<1.22$. However, to find these accurate
 values of $m_{h^\pm}$ and $m_{k^{\pm\pm}}$,  other considerations are also needed.

In the ZB model, the tiny masses of neutrino are generated at two loops, so $m_{h^\pm}$ and $m_{k^{\pm\pm}}$ cannot be very heavy \cite{test1}.
 From the experimental point of view, it is interesting to consider new scalars light enough to be produced at  the LHC.  The  theoretical arguments lead to
the fact that  the scalar masses should be a few TeVs, to avoid unnaturally large one-loop corrections to the Higgs  boson  mass which
  would cause a hierarchy problem. Therefore, these upper bounds of new scalar masses can be 2 TeVs  \cite{test}. Contacting to
   neutrino oscillation data, in the decay $k^{\pm\pm}\longrightarrow ll$, the branching ratio to $\tau\tau$ is very small in the ZB model,
   less than about $1\%$. Then, a conservative limit is $m_{k{\pm\pm}}>200$ GeV. In the ZB model, we can have the decay
   $k^{\pm\pm}\longrightarrow h^{\pm}h^{\pm}$, so $2m_{h^\pm}<m_{k^{\pm\pm}}$. Therefore, our results in Eqs. (\ref{re1}) and (\ref{re2}) are
    consistent with the above  estimation.

Recently, the experimental groups at LHC (ATLAS and CMS  Collaborations) \cite{atlas} have reported an experimental anomaly in diboson production with apparent excess in boosted jets of the $W^+ W^-, W^\pm Z$ and $ZZ$ channels at around 2 TeV invariant mass of the boson pair.

In addition, the calculation the Higgs coupling to photons (due to charged particles in the loop diagram) can be related to
 neutrino mass and $CP$ violation which are the key of matter and antimatter asymmetry. This  study  will be investigated in a future publication.

\section{CONCLUSION AND OUTLOOKS}\label{sec6}

In this paper we have investigated the EWPT in the ZB model using the high-temperature effective potential. The EWPT is strengthened
 by the new scalars to be the strongly first-order, the phase transition strength ranges from 1 to 4.15.
The new charged scalars  $h^{\pm} $ and $ k^{\pm\pm}$  are
  triggers for the first-order EWPT. Our results may be
    better  than the results in Ref. \cite{pelto}.

It is known  that if  a particle has the bigger charge,
the decay rate will be larger.  In their decay or scattering channels, we can estimate their mass, or
the parameter domain in the Higgs potential. So the charge of particle also affects the parameter domain in the Higgs potential or signatures
 of the charge of new particles ($h^\pm, k^{\pm\pm}$) are hidden in the parameter domain, which in turn can indirectly affect the EWPT.
 However, in our calculation process, we looked for the mass domain of the new particles to have a first-order phase transition.
 Then we will extract the parameter domain in the Higgs potential. If they match with values derived from scattering channels,
 our solution will be viable. Therefore, signatures of the charge may be an external condition for checking or limiting
 the parameter domain in our solution.

In addition, the EWPT can be calculated in a different way as in Ref.\cite{1101.4665, Arefe}. In order to determine $T_N$ or $T_E$,
we will examine this problem in conjunction with the $CP$-violation.

In the ZB model, the tiny mass of neutrino which can be explained at  two loop level induced by couplings between charged Higgs boson  and neutrino; and
this  can be a reason
 of the matter-antimatter asymmetry and $CP$-violation. The behavior of charged Higgs boson is also very interesting. Therefore, in the next works,
 we can  investigate the ratio $m'_{h^{\pm}/k^{\pm\pm}}$ by using neutrino data and the sphaleron rate. We will investigate the $CP$- violation
  and beyond issues of the baryon asymmetry problem through neutrino physics.

Furthermore, the sphaleron is an important process in baryogenesis and leptogenesis so we will continue to calculate and test the sphaleron
 solution in this model with the COSMOTRANSITION code \cite{cosmotran}. This code used a Bessel function for $v(r)$ but it is not flexible in
  changing the value of the wall.

With this region of self couplings in the Higgs potential, we can serve as basis for the calculation of other effects connected to  data of the LHC,
such as diphoton decay of Higgs boson, etc.

\section*{ACKNOWLEDGMENTS}

This research is funded by
 Vietnam  National Foundation for Science and Technology Development (NAFOSTED) under Grant No. 103.01-2017.356.
\\[0.3cm]

\end{document}